\documentclass[
twocolumn,
]{ceurart}

\usepackage{listings}
\usepackage{amsmath}
\usepackage{amssymb}
\usepackage{booktabs}
\usepackage{array}

\usepackage{graphicx}

\newcommand{\cival}[2]{%
  \begin{tabular}[t]{@{}c@{}}#1\\[-2pt]{\scriptsize(#2)}\end{tabular}}
\graphicspath{{figures/}}

\begin{document}

\copyrightyear{2026}
\copyrightclause{Copyright for this paper by its authors.
  Use permitted under Creative Commons License Attribution 4.0
  International (CC BY 4.0).}

\conference{RecSys in HR '26: The 6th Workshop on Recommender Systems for
  Human Resources, in conjunction with the 20th ACM Conference on
  Recommender Systems, September 28--October 2, 2026, Minneapolis, MN,
  United States.}

\title{Single-Token Expected-Value Scoring for Cold-Start Candidate Ranking
}

\author[1]{Qihang Wang}[%
email=wqihang@indeed.com,
]

\author[1]{Jinwei Tan}[%
email=jint@indeed.com,
]

\author[1]{Mengyuan Shi}[%
email=mshi@indeed.com,
]

\author[1]{Mayank Sharma}[%
email=masharma@indeed.com,
]

\author[1]{Shuai Zhao}[%
email=szhao@indeed.com,
]

\author[1]{Fuxian Li}[%
email=fuxianl@indeed.com,
]

\author[1]{Ryan Yan}[%
email=myan@indeed.com,
]

\author[1]{Alexander P. Kreuzer}[%
email=akreuzer@indeed.com,
]

\author[1]{Mohit Jain}[%
email=mjain@indeed.com,
]

\author[1]{Dheeraj Toshniwal}[%
email=dtoshniwal@indeed.com,
]

\author[1]{Manoj Seethamsetty}[%
email=manojs@indeed.com,
]

\address[1]{Indeed, Inc.}

\begin{abstract}
  AI-assisted sourcing helps streamline the initial stages of candidate
  review, reducing the administrative burden of manual screening for
  recruiters. However, deploying language models as production rankers
  remains challenging. Zero-shot Large Language Models (LLMs) may produce
  unstable, non-deterministic scores and provide limited ranking accuracy,
  while conventional deep neural rankers require large interaction logs
  that are often unavailable in cold-start settings. We use \emph{cold
  start} in a specific sense. Fitting a deep neural ranker for recruitment
  from scratch requires orders of magnitude more logged interaction data
  than a low-traffic, specialized sourcing platform produces; in our
  experience training such rankers in production, on the order of millions
  of interactions. That behavioral signal is what our setting lacks. What is
  available instead is a few hundred thousand ordinal relevance labels,
  which are small by ranker-training standards but sufficient here because a
  pretrained language model already encodes the general world knowledge the
  task depends on, allowing fine-tuning to supply domain alignment. We present
  \emph{single-token expected-value scoring}, a ranking
  primitive that casts candidate--job relevance as an ordinal
  classification over the grade tokens $\{1, \dots, 5\}$ and reads the
  relevance score as the mathematical expectation of the first-token
  probability distribution. Because the score is derived from a single
  decoding step rather than open-ended generation, it is a deterministic
  function of the model's output logits, requires no output parsing, and
  serves at low latency. To let this primitive learn the non-linear
  interdependencies of heterogeneous hiring criteria from this supervision
  alone, we fine-tune a Small Language Model (SLM) with a hybrid ordinal
  regression
  loss that combines a Mean Squared Error (MSE) term, preserving ordinal
  distance with a categorical Cross-Entropy (CE) term that sharpens
  class boundaries. The resulting framework is data-efficient, converging
  in cold-start, low-traffic regimes without the large interaction logs
  deep neural rankers demand. We assess ranking quality along two relevance
  dimensions --- Jobseeker Relevance (how well a job matches a jobseeker's
  preference) and Employer Relevance (how well a jobseeker meets an
  employer's requirements) --- using NDCG@10 and low relevance rate (the
  fraction of top results that are poor matches). Offline, our fine-tuned
  model outperforms the heuristic baseline and zero-shot LLMs. An
  end-to-end simulation shows the same direction at larger magnitude, with a
  54.2\% increase in Jobseeker Relevance NDCG@10, a 46.7\% reduction in low
  relevance rate, and consistent Employer Relevance gains; these simulation
  figures are large partly because the heuristic baseline optimizes only the
  employer side, leaving substantial jobseeker-side headroom. In a live
  online experiment it reduces employer low-relevance by 27.3\% and raises the
  employer keep rate by 7.07\%. Our work offers a
  stable and efficient ranking method that remains competitive under severe
  data constraints.
\end{abstract}


\ExplSyntaxOn
\newcommand\ceuractivateblind{ \keys_set:nn { ceur / mktitle } { blind = true } }
\ExplSyntaxOff
\makeatletter
\@ifundefined{ceurblindsubmission}{}{\ceuractivateblind}
\makeatother
\maketitle

\section{Introduction}

Hiring teams often review many resumes and profiles to decide which
candidates to contact, a process known as candidate sourcing. AI systems
are increasingly used to improve this process and reduce manual screening
\cite{aisourcing}. Because both candidate profiles and job descriptions are
composed of rich, unstructured natural language, the domain
is well-suited to Large Language Models (LLMs) and Small
Language Models (SLMs). A growing number of commercial
products and academic frameworks now leverage these generative
models to reason about the semantic alignment between candidates and
open roles \cite{llm4rec_hiring, genai_recruitment}.

However, training ranking models in this domain presents
empirical challenges. Traditional neural network-based approaches
typically require vast amounts of historical interaction data to
adequately fit the complex distributions of recruitment data
\cite{neural_ranker_data}. Emerging platforms or low-traffic products
typically lack historical interaction data, making it difficult to train
machine learning models, and handcrafted heuristic methods are used as an
alternative \cite{coldstart_heuristics}. We use \emph{cold start} in this
specific sense throughout: the constraint is the volume of
behavioral signal, not the availability of supervision. Training a deep neural
ranker from scratch on recruitment data requires interaction logs orders of
magnitude larger than a specialized, low-traffic platform accumulates,
whereas ordinal relevance labels can be generated on demand by an LLM
labeler (Section~\ref{sec:st-scoring}). A few hundred thousand such labels
remain small by ranker-training standards, and fine-tuning succeeds at that
scale only because the pretrained backbone already supplies the general
world knowledge the task depends on. Our results bear this out: for the
GPT-4.1-mini variant, gains from 20k to 50k labels are already small
(Section~\ref{sec:offline}), and the binding constraint proves to be the
capacity of the base model rather than the volume of task data
(Section~\ref{sec:analysis}).

By delegating
semantic rule evaluation and multi-dimensional criteria judgment
to LLMs, hybrid frameworks that pair these models with
traditional heuristics have become common
\cite{taxonomy_guided}. Within these pipelines, LLMs parse
unstructured text to verify specific job requirements,
connecting rigid algorithmic rules with flexible semantic
understanding.

Despite the popularity of these LLM-heuristic hybrid approaches, two
bottlenecks persist. First, the individual criteria evaluated by the
LLM and the heuristic components rarely aggregate in a simple linear
fashion, making it difficult to capture the nuanced, non-linear
interactions required for optimal candidate ranking. Second, when
relying on zero-shot LLMs, the models are frequently
relegated to binary hard-filtering of explicit constraints rather than
executing continuous, calibrated ranking \cite{finegrained_relevance}.
These zero-shot LLMs, accessed via APIs, also introduce
consistency issues; the stochastic nature of LLMs often yields volatile
results across different API calls, a vulnerability that becomes
particularly pronounced when processing large token inputs or generating
open-ended responses \cite{llm_volatility}.

In this work, we propose \emph{single-token expected-value scoring} for
cold-start candidate ranking. Our contributions are:

\begin{itemize}
\item \textbf{A deterministic ranking primitive.} We score candidate--job
  relevance as the expected value of the first-token probability
  distribution over the ordinal grade tokens $\{1, \dots, 5\}$. Computing
  the score from the first-token logits in a single decoding step makes it
  deterministic, removes output-parsing failures, and keeps inference
  latency low --- directly addressing the volatility of open-ended LLM
  generation.
\item \textbf{A data-efficient training recipe for cold start.} We
  fine-tune an SLM with a hybrid ordinal regression loss (MSE + CE) that
  enables the single-token score to capture how heterogeneous hiring
  criteria interact non-linearly. The recipe converges on LLM-generated
  relevance labels alone, without the large interaction logs that deep
  neural rankers require.
\item \textbf{Evidence from offline and online experiments.}
  Across offline ranking metrics, an end-to-end sourcing simulation, and a
  live online experiment, the method outperforms both the heuristic
  baseline and zero-shot LLMs, and we analyze where and why the gains arise
  (dual-sided relevance, ordinal integrity, and data scaling).
\end{itemize}

We validate the efficacy of our method by comparing it against both
zero-shot LLM baselines and established heuristic scoring systems.
Experimental results demonstrate that our proposed framework not only
yields better alignment and predictive accuracy in low-data regimes but
also maintains the computational stability required for real-world
deployment, where it has been successfully integrated into a production
talent sourcing platform.

\section{Background \& Problem Setting}

We first describe the automated sourcing pipeline and formalize the
ranking problem. The production recruitment system is structured as a decoupled,
three-stage pipeline designed to help recruiters navigate large candidate
pools efficiently. The pipeline begins with a retrieval phase, where an LLM
performs strategic planning based on the recruiter's explicit
requirements to formulate structured queries
\cite{llm_query_understanding}, consolidating the outputs into an
initial candidate pool at the scale of thousands of candidates. Because this initial stage leverages the
platform's legacy infrastructure, its internal mechanisms
remain outside the scope of this paper.

Following retrieval, the core candidate evaluation occurs during the
criteria judgment and scoring phase, which serves as the primary focus
of this work. In this phase, an LLM parses the employer's unstructured
job specification into distinct required and preferred criteria. The
system then evaluates the candidate profile against each individual
criterion and aggregates these granular judgments into a single relevance
score using a handcrafted heuristic. This specific method of decomposing
complex, unstructured documents into explicit, multi-dimensional
attributes for individual evaluation has been validated in recent
literature as an effective strategy to improve
alignment precision and interpretability of text-matching systems
\cite{aspect_matching}. In the final stage, the top-$N$ candidates
aggregated from multiple retrieval streams undergo global listwise
reranking, accompanied by the generation of natural language insights, a
module that is excluded from our present optimization scope.

The primary technical bottleneck of this architecture lies in the
second stage. The heuristic implicitly treats
overall candidate relevance as a linear, separable function of
independent criteria, operating under the assumption that individual
scores can simply be linearly added. In real-world talent acquisition,
however, true relevance is inherently non-linear and characterized by
complex interdependencies among qualifications, where the satisfaction of
one critical requirement may heavily outweigh or alter the significance
of others.

Formulating a robust solution to this problem presents a
behavioral-signal constraint rather than a supervisory one. While
training a standard deep neural ranker from scratch could theoretically
capture these non-linear interactions, it is infeasible in this context.
Sourcing criteria are highly dynamic, varying continuously across
different job roles and user sessions, and the low-traffic nature of
specialized recruitment platforms yields sparse interaction signals.

Relying on zero-shot LLMs to perform this aggregation introduces
substantial variance. Without task-specific alignment, these models are
inconsistent: they produce conflicting reasoning and different scores for
identical candidate contexts, failing to meet the consistency constraints of
the task. This instability is well documented. Repeated calls on unchanged
inputs yield different outputs even at low temperature
\cite{llm_volatility}, and LLM scorers are systematically sensitive to
presentation factors that carry no information about the item being judged,
such as the position an item occupies in the prompt
\cite{judge_position_bias}. Graded relevance prompting reduces but does not
remove this variance \cite{finegrained_relevance}.

We therefore reformulate the task as an ordinal classification
problem with a fine-tuned SLM. Given the parsed required and preferred
criteria alongside the unstructured candidate profile, the SLM is
optimized to map the input context directly to a discrete ordinal
relevance label within the set $\{1, 2, 3, 4, 5\}$, which subsequently
dictates the intra-job candidate ranking. We fine-tuned the model to align with complex
marketplace preferences while keeping the scoring consistency
and low latency that deployment requires.

\section{Approach}

\subsection{Heuristic Baseline}
\label{sec:baseline}

\begin{figure}[ht]
  \centering
  \includegraphics[width=\columnwidth]{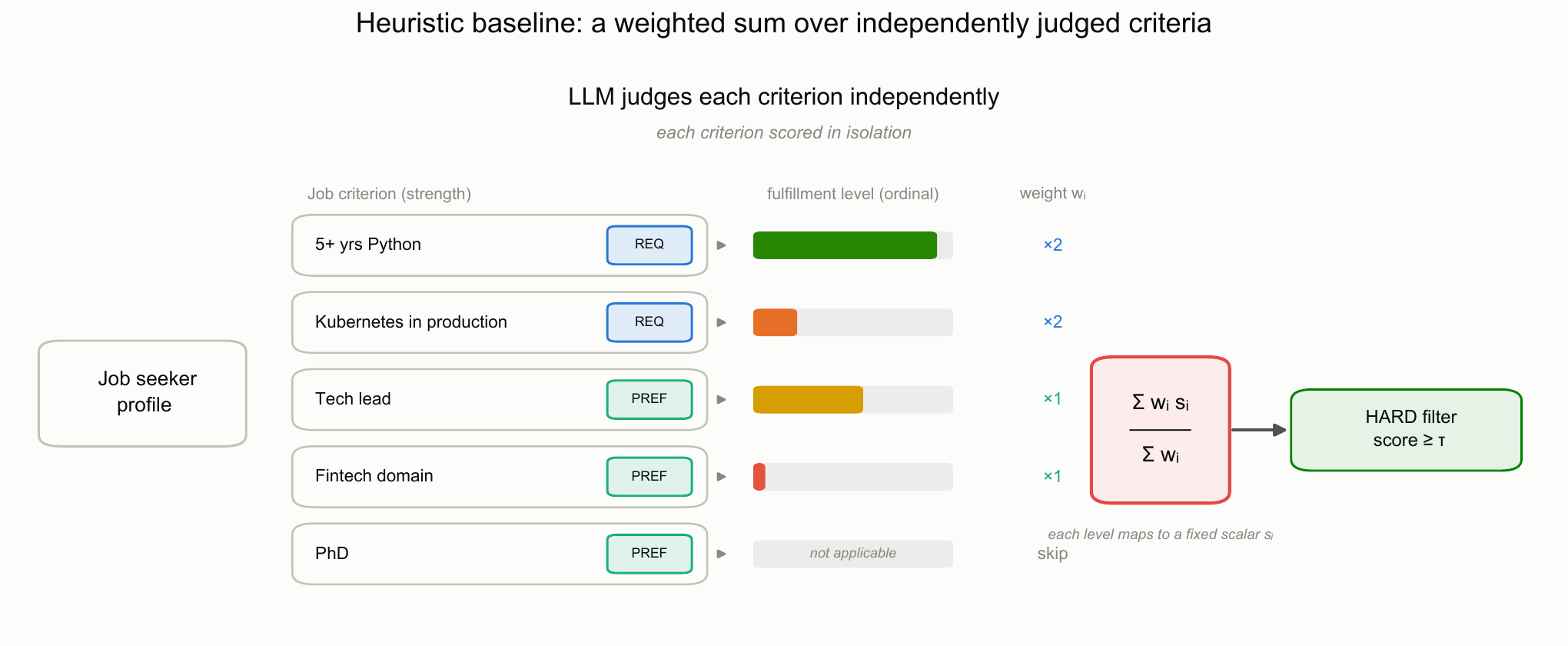}
  \caption{Heuristic baseline. A job's requirements are decomposed into
    required and preferred criteria; an LLM judges the candidate against
    each criterion independently, and the per-criterion scores are
    aggregated into a single relevance score by a weighted linear sum
    before thresholding.}
  \label{fig:baseline}
\end{figure}

During onboarding, each job's requirements are decomposed into a set of
required and preferred criteria, either provided by the employer or
generated from the job description. For a given candidate, an LLM judges
each criterion independently, assigning an ordinal fulfillment level
(ranging from explicitly satisfied to not met) based on the evidence in
the candidate profile. Each level is mapped to a fixed scalar, and the
per-criterion scores are combined into a single relevance score via a
weighted linear sum, with required and preferred criteria weighted
differently. Candidates are then thresholded on this aggregate score to
form the set passed to downstream listwise reranking. This design
embodies the additive separability assumption we challenge in this work:
overall relevance is treated as a linear sum of independently scored
criteria, with no mechanism to model interactions between them
(Figure~\ref{fig:baseline}).

\subsection{Single-Token Expected-Value Scoring}
\label{sec:st-scoring}

We follow the labeling framework of \citet{ordinal_relevance}. For each
job--jobseeker pair, an LLM labeler produces two complementary relevance
signals: \emph{Employer Relevance}, which measures how well the jobseeker
satisfies the employer's requirements, and \emph{Jobseeker Relevance},
which measures how well the job aligns with the jobseeker's preferences or
likely interests. A fixed, deterministic policy maps the employer-side
verdict and the jobseeker-side grade onto a single overall-relevance
training label $y \in \{1, 2, 3, 4, 5\}$: 5 when both sides are the best fit
for one another and 1 when either side is a bad match. Observed interaction
outcomes take precedence where available. The policy is applied identically
across the train, validation, and test splits, so no label drift is
introduced across the time-based split of Section~\ref{sec:offline}.

Here we use a pointwise approach, where the model produces a single score
based on a pair of job and jobseeker. With the above labels, it is common
to constrain the model to emit a single grade token and use the token
probability as the ranking score. Our formulation is motivated by
\citet{finegrained_relevance}, who show that prompting an LLM ranker over
a fine-grained ordinal label set and reading the \emph{expected value} of
the resulting distribution is substantially more effective than the
conventional binary yes/no relevance probability, as the graded scale
exposes finer distinctions that a single relevant-token probability
cannot. Building on this idea, instead of taking the most likely grade we
compute the \emph{Expected Relevance} (ER): the expectation of the ordinal
grade under the model's predicted distribution over the grade tokens. We
formalize this expected-value score in Eq.~\eqref{eq:expected} and use it
both as the inference-time ranking score and as the regression target of
our training loss.

Our departure from \citet{finegrained_relevance} is that they apply the
expected-value score in a purely zero-shot prompting setting, whereas we
turn it into a learnable training objective: we fine-tune the model with a
hybrid ordinal regression loss (detailed below) that directly optimizes
this expected-value score, combining a regression term with a
classification term rather than relying on the pretrained distribution
alone.

We experimented with both LLMs and SLMs:
\begin{itemize}
\item For the LLM, we take the first-token distribution, keep only the
  probabilities of the grade tokens 1--5 (renormalized), and compute
  their expected value. Standard commercial APIs support \texttt{logprobs}
  and \texttt{logit\_bias}, allowing us to constrain the output to a single
  grade token and read its probability distribution.
\item For the SLM, we do not need to specify the output token via
  \texttt{logit\_bias}; we still calculate the final value using the
  expected-value method.
\end{itemize}

To effectively capture the inherent numerical and ordinal relationships
in score-based token classification tasks (e.g., predicting ratings from
1 to 5), we utilize a hybrid \emph{Ordinal Regression Loss}. The
objective function consists of a joint optimization framework combining a
Mean Squared Error (MSE) regression component with a standard
Cross-Entropy (CE) classification penalty.

Let $N$ denote the number of valid target samples in a batch, and let $K$
represent the total number of ordinal classes. For each sample $i \in
\{1, 2, \dots, N\}$:
\begin{itemize}
\item $\mathbf{z}_i = [z_{i1}, z_{i2}, \dots, z_{iK}]^\top \in
  \mathbb{R}^K$ represents the logit vector extracted from the sequence
  position immediately preceding the target token.
\item $\mathbf{v} = [v_1, v_2, \dots, v_K]^\top \in \mathbb{R}^K$ denotes
  the static vector mapping continuous score values to each class (e.g.,
  $\mathbf{v} = [1.0, 2.0, \dots, 5.0]^\top$).
\item $y_i \in \{1, 2, \dots, K\}$ is the ground-truth class index.
\item $s_i = v_{y_i}$ represents the true scalar score value
  corresponding to the ground-truth class.
\end{itemize}

The categorical probability $p_{ik}$ that the $i$-th sample belongs to
class $k$ is derived via the softmax operation over the target logits:
\begin{equation}
  p_{ik} = \frac{\exp(z_{ik})}{\sum_{j=1}^K \exp(z_{ij})} .
\end{equation}

Using the predicted class probabilities as weights, we compute the
continuous \emph{expected score} $\hat{s}_i$ for sample $i$ as
\begin{equation}
  \label{eq:expected}
  \hat{s}_i = \sum_{k=1}^K p_{ik} \cdot v_k .
\end{equation}
This expected score is exactly the Expected Relevance (ER) introduced
above: it serves as the regression target in the loss below and, at
inference time, as the ranking score. With the linear value map
$v_k = k$, it reduces to $\hat{s}_i = \sum_{k=1}^K k\, p_{ik}$.

\paragraph{Mean Squared Error (MSE) Loss.}
The regression component minimizes the squared Euclidean distance between
the continuous expected score and the true continuous score. This
explicitly forces the network to respect the ordinal distance between
distinct rating levels:
\begin{equation}
  \mathcal{L}_{\text{MSE}} = \frac{1}{N} \sum_{i=1}^N (\hat{s}_i - s_i)^2 .
\end{equation}

\paragraph{Cross-Entropy (CE) Loss.}
To maintain stable class discrimination and accelerate alignment, a
standard categorical cross-entropy loss is simultaneously optimized:
\begin{equation}
  \mathcal{L}_{\text{CE}} = -\frac{1}{N} \sum_{i=1}^N \log p_{i, y_i} .
\end{equation}

The final joint loss function $\mathcal{L}_{\text{total}}$ is formalized
as a weighted sum of the two complementary objectives:
\begin{equation}
  \label{eq:total}
  \mathcal{L}_{\text{total}} = w_{\text{MSE}} \cdot \mathcal{L}_{\text{MSE}}
  + w_{\text{CE}} \cdot \mathcal{L}_{\text{CE}} ,
\end{equation}
where $w_{\text{MSE}}$ and $w_{\text{CE}}$ are hyper-parameters
controlling the regularization trade-off between the expected-value
distance and strict categorical classification performance. Our deployed
Qwen3-8B SFT model uses $w_{\text{MSE}} = 0.6$ and $w_{\text{CE}} = 0.4$;
we ablate this choice against the two single-objective extremes in
Section~\ref{sec:loss-ablation}.

\subsection{Fine-Tuning Configuration}
\label{sec:finetune}

We fine-tune the Qwen3-8B backbone with Low-Rank Adaptation (LoRA),
freezing the base weights and training only lightweight adapters on all
attention and MLP projection matrices, which keeps adaptation
parameter- and memory-efficient. We optimize the hybrid ordinal regression
loss of Eq.~\eqref{eq:total} with the paged AdamW 8-bit optimizer under a
linear learning-rate schedule, training across 8 GPUs with DeepSpeed
ZeRO stage~2. Table~\ref{tab:finetune} summarizes the configuration.

\begin{table}[ht]
  \centering
  \caption{Fine-tuning configuration for the Qwen3-8B SLM ranker.}
  \label{tab:finetune}
  \footnotesize
  \setlength{\tabcolsep}{4pt}
  \begin{tabular}{@{}l p{0.60\columnwidth}@{}}
    \toprule
    Component & Configuration \\
    \midrule
    Backbone          & Qwen3-8B \\
    Adaptation        & LoRA ($r{=}16$, $\alpha{=}32$, dropout $0.05$) on all attention and MLP projections \\
    Optimizer         & paged AdamW 8-bit, weight decay $0.01$ \\
    Learning rate     & $1{\times}10^{-5}$, linear schedule with warmup \\
    Effective batch   & 512 ($4$ per device $\times\,16$ grad.\ accum.\ $\times\,8$ GPUs) \\
    Hardware          & 8 GPUs, DeepSpeed ZeRO-2 \\
    \bottomrule
  \end{tabular}
\end{table}

\section{Offline Experiments and Results}
\label{sec:offline}

In this section, we conduct an offline evaluation to assess the efficacy
of our proposed learned ordinal relevance framework. We evaluate on a
dataset of job--jobseeker pairs carrying the overall relevance labels of
Section~\ref{sec:st-scoring}, retaining the full candidate list from the
retrieval stage for each job. We use a time-based split in which training
and validation precede the held-out test period, mirroring how production
systems are deployed on future, unseen traffic and avoiding temporal
leakage. Table~\ref{tab:systems} lists every system we compare and the data it
was trained on; we use these names throughout.

\begin{table}[t]
  \caption{Systems compared in this paper. All fine-tuned variants optimize
    the hybrid ordinal regression loss of Eq.~\eqref{eq:total}. Every system
    is scored by the expected-value read-out (``ER'') of
    Eq.~\eqref{eq:expected} over the grade tokens, except the heuristic
    baseline, which uses a weighted linear sum, and GPT-5.4, which is scored
    from open-ended generation ($\dagger$). The peak-score alternative $p(5)$
    is evaluated as an ablation in Section~\ref{sec:peak-vs-er} rather than
    as a separate system.}
  \label{tab:systems}
  \footnotesize
  \setlength{\tabcolsep}{3pt}
  \begin{tabular}{@{}>{\raggedright\arraybackslash}p{0.315\columnwidth}>{\raggedright\arraybackslash}p{0.165\columnwidth}>{\raggedright\arraybackslash}p{0.245\columnwidth}>{\raggedright\arraybackslash}p{0.175\columnwidth}@{}}
    \toprule
    \textbf{System} & \textbf{Base model} & \textbf{Adaptation / data} &
    \textbf{Used in} \\
    \midrule
    Heuristic baseline & --- & hand-tuned weights & 4.2, 4.5, 5.1, 5.2 \\
    GPT-5.4 (zero-shot)$^{\dagger}$ & GPT-5.4 & none & 4.2 \\
    GPT-4.1-mini (zero-shot) & GPT-4.1-mini & none & 4.2, 4.3 \\
    GPT-4.1-mini-ft (20k) & GPT-4.1-mini & SFT, 20k pairs & 4.2, 4.3 \\
    GPT-4.1-mini-ft (50k) & GPT-4.1-mini & SFT, 50k pairs & 4.3, 4.4, 5.1, 7.2 \\
    Qwen3-8B-ft (50k) & Qwen3-8B & LoRA SFT, 50k pairs & 4.4, 7.2 \\
    \textbf{Qwen3-8B SFT (200k)} \emph{(deployed)}
      & Qwen3-8B & LoRA SFT, 200k pairs & 4.4--4.7, 5.1, 5.2, 6 \\
    \bottomrule
  \end{tabular}

  \vspace{2pt}
  {\footnotesize $^{\dagger}$GPT-5.4 does not expose logit\_bias or
  constrained logprobs, so it cannot be scored by the expected-value
  read-out and is instead scored from open-ended generation. Its numbers are
  therefore not directly comparable to the other zero-shot entry.}
\end{table}

\subsection{Experimental Setup and Evaluation Metrics}
\label{sec:setup}

To evaluate the ranking performance of the candidate sourcing models, we
adopt the following standard metrics:

\begin{itemize}
\item \textbf{Spearman's Rank Correlation ($\rho_{\text{global}}$):}
  Measures the global monotonic relationship between the predicted
  rankings and the reference overall relevance labels.
\item \textbf{Normalized Discounted Cumulative Gain (NDCG@$k$, $k \in
  \{5, 10, 20\}$):} Evaluates the position-aware relevance quality of
  the top-$k$ ranked candidates.
\item \textbf{Recall@$k$ ($k \in \{5, 10, 20\}$):} Measures the
  proportion of highly relevant candidates successfully retrieved within
  the top-$k$ positions.
\item \textbf{High-relevance Rate@$k$:} Proportion of the top-$k$
  candidates that are good matches. We compute this for both
  jobseeker-side and employer-side relevance labels; each label is defined
  on a 1--5 ordinal scale, where scores of 4 and 5 are classified as
  highly relevant.
\item \textbf{Low-relevance Rate@$k$:} Proportion of the top-$k$
  candidates that are poor matches. We compute this for both
  jobseeker-side and employer-side relevance labels; each label is defined
  on a 1--5 ordinal scale, where scores of 1 and 2 are classified as
  low-relevance.
\item \textbf{Pairwise Accuracy:} Evaluates the model's capability to
  correctly identify the superior candidate within a given pair across
  distinct ordinal overall relevance classes.
\end{itemize}

\subsection{Performance Comparison: Heuristic Baseline vs. Zero-Shot vs. Fine-Tuning}

\begin{figure}[ht]
  \centering
  \includegraphics[width=\columnwidth]{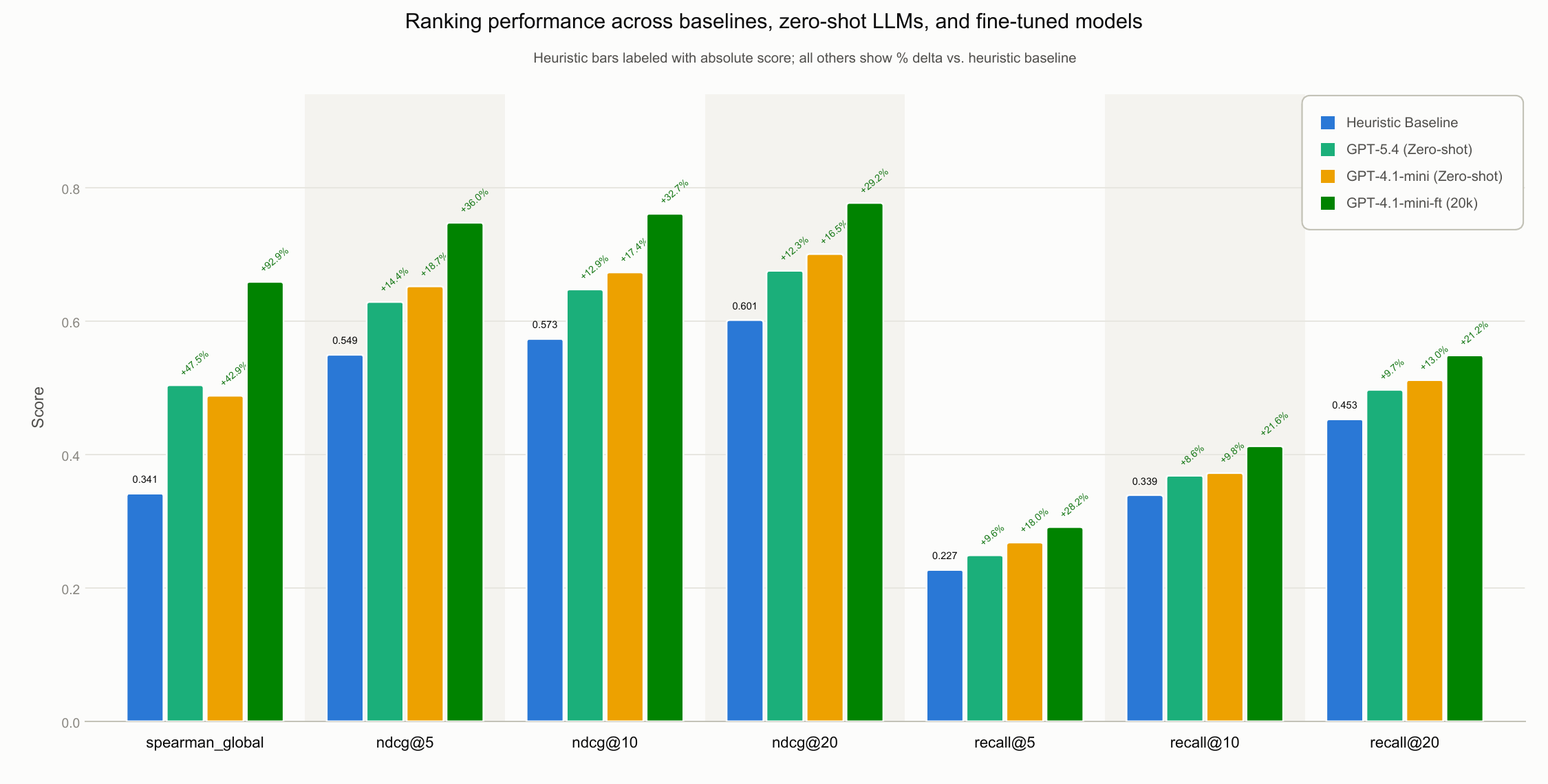}
  \caption{Performance comparison across baselines, zero-shot
    LLMs, and fine-tuned models. Metrics: Spearman's $\rho$,
    NDCG@$\{5,10,20\}$, and Recall@$\{5,10,20\}$. Zero-shot LLMs already
    surpass the heuristic, and fine-tuning adds a further consistent gain
    across every metric (e.g., $+34.9\%$ in global Spearman over
    GPT-4.1-mini zero-shot). GPT-5.4 is scored from open-ended generation
    rather than by the expected-value read-out, as its API exposes neither
    \texttt{logit\_bias} nor constrained \texttt{logprobs}; its bars are
    therefore not a like-for-like measure of model capability.}
  \label{fig:perf-compare}
\end{figure}

In Figure~\ref{fig:perf-compare}, zero-shot LLM configurations
consistently outperform the heuristic baseline. This confirms that
leveraging the pre-trained knowledge of generative models yields
better text alignment than rigid, additive scoring methods.

Our fine-tuned model (GPT-4.1-mini-ft-20k) further improves over the
zero-shot alternatives, achieving a 34.9\%
improvement in global Spearman correlation over GPT-4.1-mini (Zero-shot).

\paragraph{Note on Model Constraints.}
GPT-5.4 (Zero-shot) displays a slightly lower NDCG than
GPT-4.1-mini (Zero-shot). This disparity is primarily attributed to API
limitations: GPT-5.4 lacks support for logit\_bias and constrained
logprobs parameterization. Consequently, it is restricted to open-ended
token generation, preventing the precise mathematical-expectation
modeling utilized by GPT-4.1 and introducing output volatility.

\subsection{Impact of Fine-Tuning Data Scale}

\begin{figure}[ht]
  \centering
  \includegraphics[width=\columnwidth]{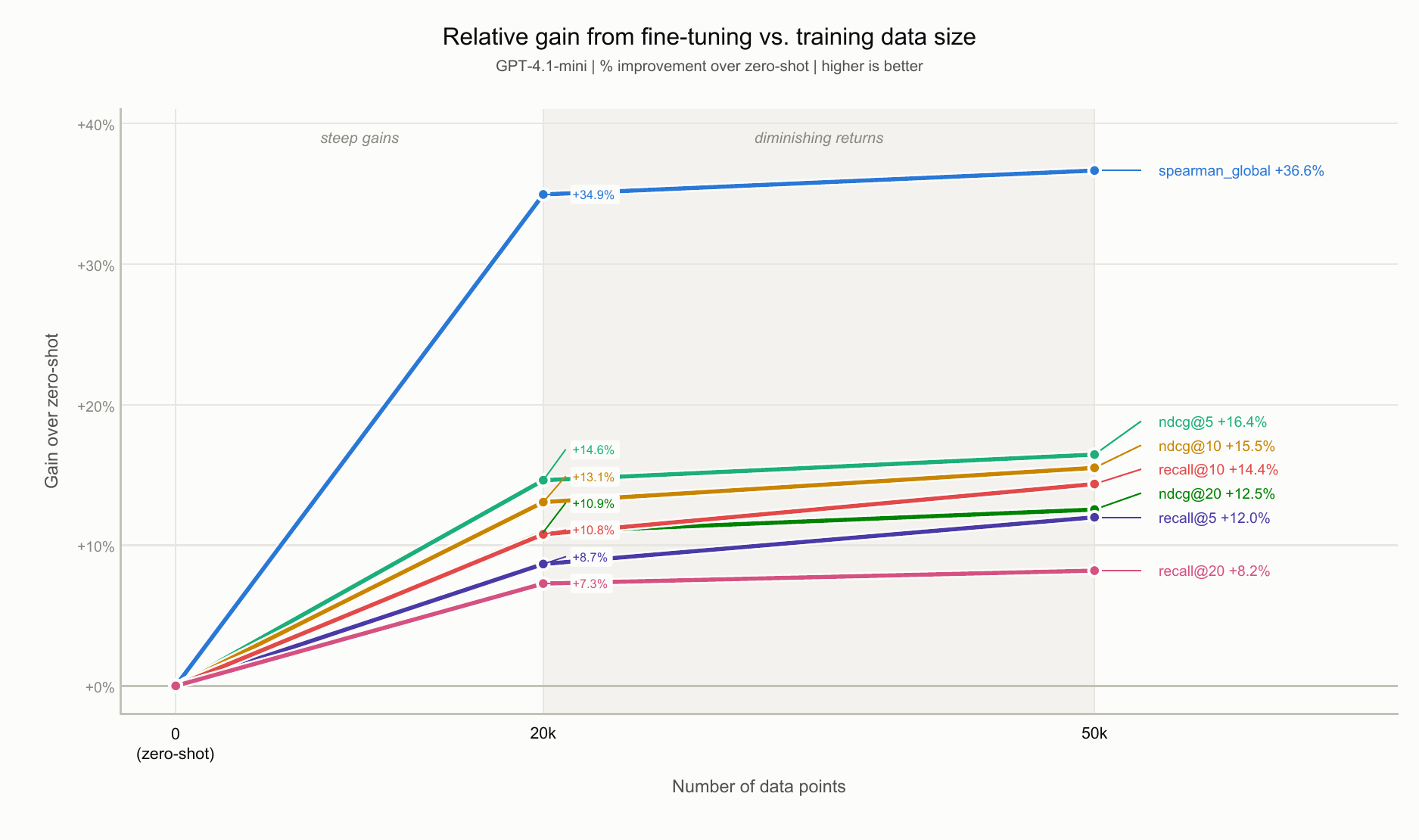}
  \caption{Performance dynamics for GPT-4.1-mini under varying fine-tuning
    data sizes. Most of this backbone's ranking quality is reached by
    ${\sim}$20k labels, with only incremental gains from 20k to 50k.}
  \label{fig:data-scale}
\end{figure}

To explore the data efficiency and scaling behaviors of our framework, we
evaluate the ranking performance across varying sizes of the
annotated training dataset.

The results in Figure~\ref{fig:data-scale} indicate that expanding the
fine-tuning corpus from 20k to 50k instances delivers gains
across all ranking and recall dimensions. However, the magnitude of
improvement is considerably less pronounced than the transition
from zero-shot to supervised fine-tuning. Under the evaluated
configuration, this suggests that meaningful task adaptation can be
achieved with tens of thousands of relevance labels, even in the absence
of large-scale behavioral interaction logs; for this backbone, additional
labeled data continues to improve ranking quality but with diminishing
marginal gains. Section~\ref{sec:analysis} shows that the point at which
returns diminish is itself backbone-dependent.

\subsection{Proprietary LLM vs. On-Premise SLM Deployment Trade-offs}

For real-world deployment, operational variables such as serving costs,
API deprecation risks, and inference latency are critical. We therefore
evaluate a locally hosted open-source alternative, Qwen3-8B, against the
proprietary GPT-4.1-mini-ft. To separate the effect of the backbone from
the effect of training-set size, we first fine-tune both on the same 50k
pairs.

\begin{figure}[ht]
  \centering
  \includegraphics[width=\columnwidth]{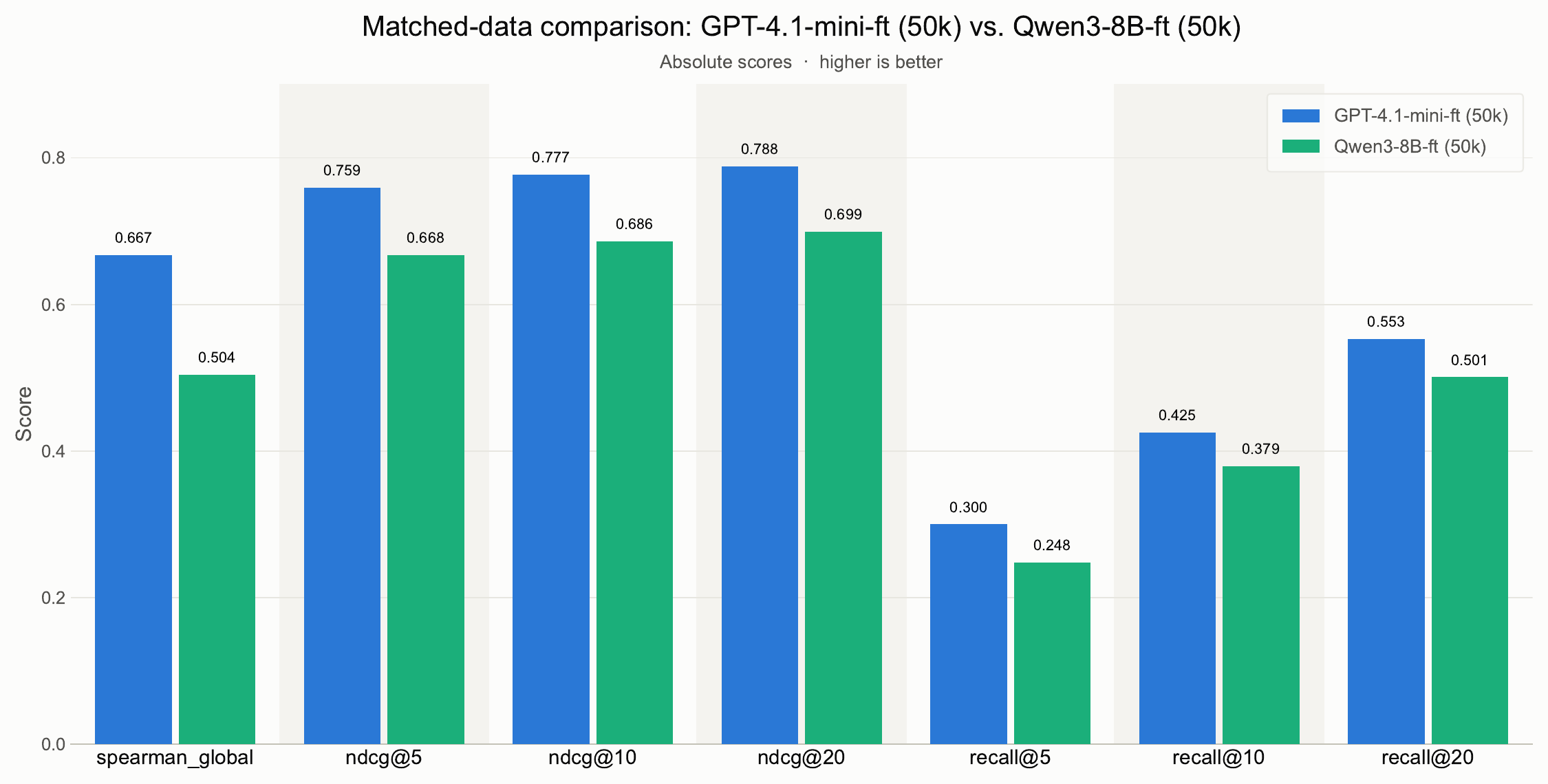}
  \caption{Matched-data comparison of the open-source SLM and the
    fine-tuned proprietary LLM, both trained on 50k pairs. Holding the
    training set fixed isolates the effect of the backbone:
    GPT-4.1-mini-ft leads on every metric, by $+10.4\%$ to $+32.3\%$, and
    most widely on global Spearman ($0.667$ vs.\ $0.504$).}
  \label{fig:slm-vs-llm}
\end{figure}

As shown in Figure~\ref{fig:slm-vs-llm}, at matched training-set size
GPT-4.1-mini-ft leads Qwen3-8B-ft on every metric, by $+10.4\%$ to
$+32.3\%$, with the widest margin on global Spearman. Because both models
see identical supervision, the gap is attributable to the backbone:
differences in pre-training scale and composition, parameter count, and
architecture. Scaling Qwen3-8B to 200k pairs recovers much of the
difference, improving on its own 50k configuration by $+5.8\%$ to
$+21.2\%$ (Table~\ref{tab:loss-ablation}, hybrid column), but leaves it
$4.2\%$ to $8.4\%$ behind GPT-4.1-mini-ft at 50k. Additional task data
narrows the gap without closing it.

However, from an engineering and production standpoint, Qwen3-8B SFT
maintains solid metrics that clearly outperform the legacy
heuristic. Given its advantages in cost efficiency, full control
over serving latency, and mitigation of upstream commercial API
deprecation risks, Qwen3-8B SFT was the preferred choice for production
deployment.

\subsection{Pairwise Accuracy}

To examine the fine-grained discrimination of our deployment
model, we analyze the pairwise accuracy across different combinations of
the 1-to-5 ordinal relevance labels.

\begin{figure}[ht]
  \centering
  \includegraphics[width=\columnwidth]{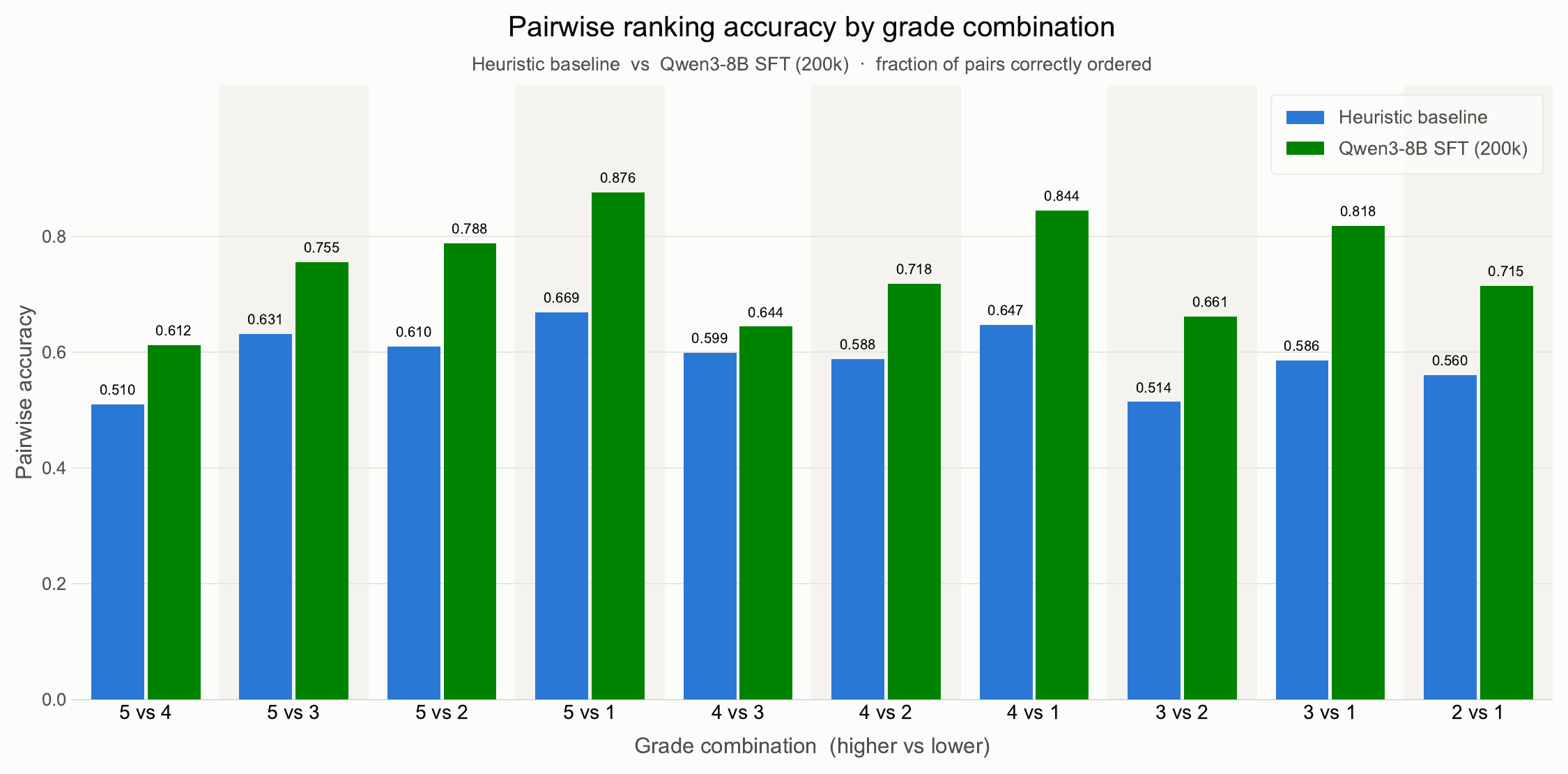}
  \caption{Pairwise accuracy breakdowns across ordinal score
    combinations. The model separates distant grades sharply (87.64\% on 5
    vs 1) while adjacent tiers (5 vs 4, 4 vs 3) remain hardest; on adjacent
    pairs it stays above ${\sim}$61\% vs.\ the heuristic's near-random
    ${\sim}$51\%, showing the ordinal loss preserves grade order.}
  \label{fig:pairwise}
\end{figure}

Figure~\ref{fig:pairwise} shows that Qwen3-8B SFT achieves higher pairwise
ranking accuracy than the heuristic baseline across all grade
combinations. Performance is strongest for widely separated grades; for
example, accuracy reaches 87.64\% for the 5-versus-1 comparison. In
contrast, adjacent grades, such as 5 versus 4 and 4 versus 3, remain more
difficult to distinguish because the differences between neighboring
relevance levels are more subtle. Overall, this pattern suggests that the
fine-tuned model better captures the ordinal structure of the relevance
labels, while fine-grained distinctions between adjacent grades remain
challenging.

\subsection{Peak Score vs.\ Expected Value}
\label{sec:peak-vs-er}

We consider two formulas for collapsing the model's predicted
distribution over the grade tokens into a single ranking score. The first, which we call the
\emph{Peak Score}, discards the full distribution and ranks by the
probability mass assigned to the top grade $p(5)$ --- effectively asking
how confident the model is that a candidate is a top-tier match. The
second is the \emph{Expected Relevance} (ER) of
Eq.~\eqref{eq:expected}, which aggregates the entire $1$--$5$ distribution
into a single continuous expectation. To decide which formula to serve, we
recompute the pairwise accuracy under both scores across the same ordinal
combinations.

\begin{figure}[ht]
  \centering
  \includegraphics[width=\columnwidth]{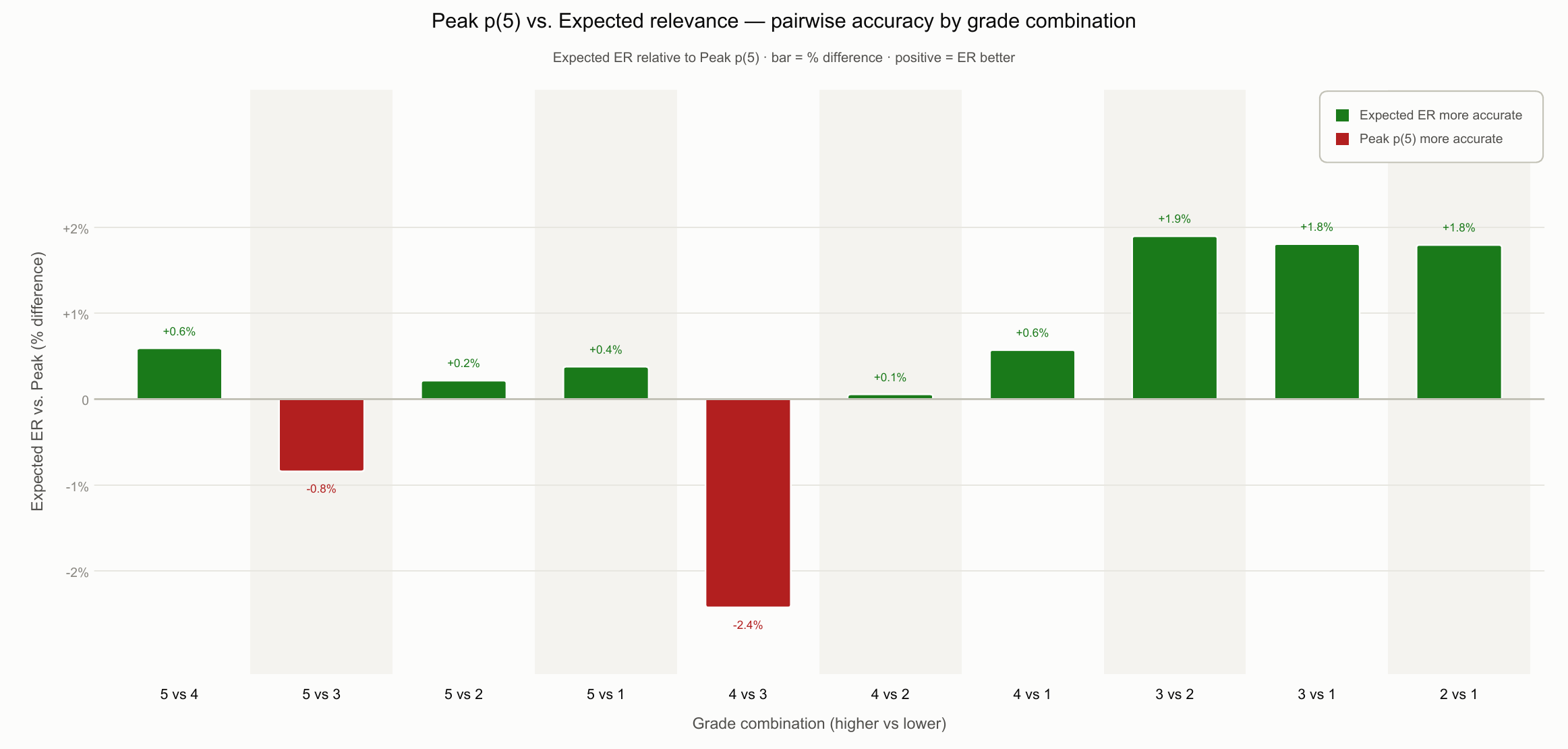}
  \caption{Pairwise accuracy difference under Peak Score $p(5)$
    versus Expected Relevance (ER) scoring. ER is on par or better on most
    pairs and wins overall ($0.7353$ vs.\ $0.7306$), with its largest gains
    on harder low/mid-band separations; we therefore serve ER.}
  \label{fig:peak-vs-er}
\end{figure}

As shown in Figure~\ref{fig:peak-vs-er}, ER edges out the Peak Score
overall ($0.7353$ vs.\ $0.7306$, a relative gain of $+0.64\%$) and wins
the majority of individual combinations. The Peak Score holds a narrow
advantage only on two pairs (\textbf{5 vs 3} and
\textbf{4 vs 3}), whereas ER delivers its clearest gains on the harder
low- and mid-band separations (e.g., \textbf{3 vs 1} at $+1.81\%$ and
\textbf{2 vs 1} at $+1.79\%$),
where collapsing the distribution to a single top-grade probability
throws away discriminative signal. Because ER performs better overall and
on eight of the ten grade-pair comparisons, while preserving information
from the full ordinal distribution, we adopt Expected Relevance as the
production ranking score. The accuracy figures reported in
Figure~\ref{fig:pairwise} correspond to this ER scoring.

\subsection{Loss-Weighting Ablation}
\label{sec:loss-ablation}

The hybrid objective in Eq.~\eqref{eq:total} blends an ordinal-distance
term (MSE) with a class-discrimination term (CE). To understand the
contribution of each component, we compare our production weighting
against the two single-objective extremes, holding the model, training
data (200k), and all other hyper-parameters fixed:
\begin{itemize}
\item \textbf{MSE-only} ($w_{\text{MSE}} = 1.0$, $w_{\text{CE}} = 0.0$):
  optimizes solely for expected-value distance, discarding the explicit
  categorical objective.
\item \textbf{CE-only} ($w_{\text{MSE}} = 0.0$, $w_{\text{CE}} = 1.0$):
  optimizes solely for grade classification, discarding the ordinal
  distance penalty.
\item \textbf{Hybrid (ours)} ($w_{\text{MSE}} = 0.6$,
  $w_{\text{CE}} = 0.4$): the deployed configuration.
\end{itemize}

To characterize the trade-off from both a classification and a ranking
angle, we augment the ranking metrics of Section~\ref{sec:setup} with three
grade-level measures:
\begin{itemize}
\item \textbf{Accuracy:} the exact-match rate --- the fraction of
  candidates whose predicted grade equals the ground-truth ordinal label.
\item \textbf{Adjacent Accuracy:} a relaxed variant that counts a
  prediction as correct when it falls within one level ($\pm 1$) of the
  ground-truth grade, tolerating near-miss ordinal errors.
\item \textbf{Mean Absolute Error (MAE):} the average absolute difference
  between the predicted and true grades (lower is better), a
  magnitude-aware complement to the rank-based metrics.
\end{itemize}

\begin{table}[t]
  \centering
  \caption{Loss-weighting ablation for Qwen3-8B SFT (200k), with $95\%$
    confidence intervals in parentheses; for MAE lower is better. No best
    value is marked, because the intervals overlap on all seven ranking
    metrics. Only exact-match accuracy and adjacent accuracy separate, and
    both isolate MSE-only.}
  \label{tab:loss-ablation}
  \footnotesize
  \setlength{\tabcolsep}{3pt}
  \begin{tabular}{@{}lccc@{}}
    \toprule
           & MSE-only & CE-only & Hybrid (ours) \\
    Metric & (1.0 / 0.0) & (0.0 / 1.0) & (0.6 / 0.4) \\
    \midrule
    accuracy             & \cival{0.3420}{0.3180, 0.3665} & \cival{0.5044}{0.4817, 0.5264} & \cival{0.5083}{0.4863, 0.5299} \\
    adjacent accuracy    & \cival{0.7906}{0.7714, 0.8098} & \cival{0.7414}{0.7211, 0.7614} & \cival{0.7556}{0.7364, 0.7745} \\
    mae ($\downarrow$)   & \cival{0.9375}{0.8947, 0.9805} & \cival{0.9602}{0.9181, 1.0017} & \cival{0.9359}{0.8933, 0.9786} \\
    \midrule
    spearman\_global     & \cival{0.6097}{0.5759, 0.6413} & \cival{0.5963}{0.5593, 0.6312} & \cival{0.6106}{0.5754, 0.6435} \\
    ndcg@5               & \cival{0.7187}{0.6846, 0.7519} & \cival{0.7097}{0.6754, 0.7442} & \cival{0.7175}{0.6834, 0.7511} \\
    ndcg@10              & \cival{0.7349}{0.7059, 0.7641} & \cival{0.7301}{0.7009, 0.7594} & \cival{0.7324}{0.7023, 0.7619} \\
    ndcg@20              & \cival{0.7472}{0.7218, 0.7721} & \cival{0.7437}{0.7181, 0.7691} & \cival{0.7479}{0.7226, 0.7727} \\
    recall@5             & \cival{0.2803}{0.2368, 0.3275} & \cival{0.2668}{0.2246, 0.3129} & \cival{0.2830}{0.2389, 0.3303} \\
    recall@10            & \cival{0.4076}{0.3575, 0.4607} & \cival{0.4016}{0.3521, 0.4547} & \cival{0.4071}{0.3571, 0.4602} \\
    recall@20            & \cival{0.5286}{0.4788, 0.5789} & \cival{0.5267}{0.4766, 0.5768} & \cival{0.5298}{0.4804, 0.5800} \\
    \bottomrule
  \end{tabular}
\end{table}

Table~\ref{tab:loss-ablation} isolates the effect of each loss term. The
ranking differences among the three objectives are small relative to their
uncertainty. Across all NDCG and Recall cutoffs the $95\%$ confidence
intervals substantially overlap, providing no clear evidence that one loss
consistently dominates the others in ranking quality. The point estimates do
show a mild and consistent advantage for the hybrid objective on recall: it
achieves the highest Recall@5 and Recall@20, while remaining effectively tied
with MSE-only at Recall@10.

The more pronounced differences appear in grade-level behavior. MSE-only
achieves strong adjacent accuracy but suffers a substantial drop in
exact-match accuracy, suggesting that optimizing ordinal distance alone tends
to produce predictions close to, but not exactly at, the target grade.
CE-only restores exact-match accuracy, but yields less favorable MAE and
Spearman results. The hybrid objective provides the most balanced operating
point: it matches CE-only on exact classification accuracy while retaining
ranking and recall performance comparable to MSE-only.

Because our downstream application places particular emphasis on recall, and
the hybrid objective provides the strongest or near-strongest recall point
estimates without the exact-classification weakness of MSE-only, we retain
$w_{\text{MSE}} = 0.6$ and $w_{\text{CE}} = 0.4$ for production. The
uncertainty estimates suggest that this choice should be interpreted as a
robustness and trade-off decision rather than as evidence of a statistically
significant ranking improvement.

\section{Simulation and Online Experiment Results}

\subsection{End-to-End Simulation}

Beyond the offline metrics, we ran an end-to-end simulation that
reproduces the full sourcing-agent workflow, letting us observe how the
ranker behaves in context.

In this evaluation, candidate pools are dynamically generated via the
upstream retrieval layer and capped at a maximum of $N = 30$ candidates
per job post to match real-world screening constraints. We deploy our
on-premise Qwen3-8B SFT (200k) model alongside the proprietary
GPT-4.1-mini-ft (50k) model to serve inference within the scoring stage.

\paragraph{Experimental Control.}
In the production environment's legacy architecture, candidate sequences
are passed to a downstream \emph{listwise reflection} module, which
applies LLM-based global heuristics and often permutes the final
sequence. However, listwise reranking can introduce stochastic noise and
structural distortions that mask the true sorting accuracy of the
underlying point-wise ranker. To isolate the direct impact of our learned
ordinal relevance and ensure a rigorous, fair empirical comparison, we
bypassed the listwise ranking layer. The final ranking presentation
explicitly preserves the deterministic order derived from our continuous
expected-value scores.

\paragraph{End-to-End Simulation Metrics.}
We evaluate the simulation results across two primary dimensions:
Employer Relevance (alignment with job descriptions) and Jobseeker
Relevance (alignment with candidate preferences and historical match
success). Performance gains are calculated as relative improvements over
the legacy heuristic baseline.

\begin{table}[ht]
  \centering
  \caption{Relative performance lift (\%) in the end-to-end simulated
    sourcing pipeline.}
  \label{tab:sim-lift}
  \footnotesize
  \setlength{\tabcolsep}{4pt}
  \begin{tabular}{@{}p{0.50\columnwidth}cc@{}}
    \toprule
    Evaluation Dimension \& Metric & Qwen3-8B & GPT-4.1-mini \\
                                   & SFT (200k) & ft (50k) \\
    \midrule
    Employer Relevance NDCG@10                   & $+3.1\%$  & $+5.9\%$ \\
    Employer Relevance NDCG@20                   & $+2.6\%$  & $+4.2\%$ \\
    Employer Relevance High Relevance Rate@20    & $+4.5\%$  & $+6.2\%$ \\
    Employer Relevance Low Relevance Rate@20     & $-23.8\%$ & $-27.3\%$ \\
    Jobseeker Relevance NDCG@10                  & $+54.2\%$ & $+63.7\%$ \\
    Jobseeker Relevance NDCG@20                  & $+36.7\%$ & $+42.4\%$ \\
    Jobseeker Relevance High Relevance Rate@20   & $+32.7\%$ & $+37.7\%$ \\
    Jobseeker Relevance Low Relevance Rate@20    & $-46.7\%$ & $-56.2\%$ \\
    \bottomrule
  \end{tabular}
\end{table}

The legacy heuristic baseline is, by construction, a one-sided ``employer
hard-requirements checklist'' hand-crafted from the job description (JD).
It therefore already encodes employer-side preferences reasonably well and
sits near a local optimum on that dimension, while leaving jobseeker-side
alignment essentially unoptimized. The simulation lifts follow this
asymmetry directly: Employer Relevance improves moderately (NDCG@10 of
$+3.1\%$ / $+5.9\%$), whereas Jobseeker Relevance, which the baseline never
targeted, rises far more ($+54.2\%$ / $+63.7\%$). The large jobseeker-side
gains therefore measure headroom in the baseline rather than an
order-of-magnitude absolute advantage for the ranker.

\subsection{Online Experiment}
\label{sec:online}

To evaluate the real-world utility of the proposed ranker, we conducted a
live online evaluation. We randomized at the employer level rather than the
account level: an employer may operate several accounts, so randomizing by
employer keeps the ranking experience consistent across all of an employer's
accounts and prevents any single employer from being exposed to both arms.
\begin{enumerate}
\item \textbf{Control:} the production heuristic baseline of
  Section~\ref{sec:baseline} (per-criterion LLM judgments aggregated by a
  hand-tuned weighted sum).
\item \textbf{Treatment:} the single-token expected-value score from the
  fine-tuned SLM (Section~\ref{sec:st-scoring}).
\end{enumerate}

\begin{table}[ht]
  \centering
  \caption{Online A/B results: relative lift of treatment
    (fine-tuned SLM) over control (heuristic), with 95\% confidence
    intervals in parentheses.}
  \label{tab:online}
  \setlength{\tabcolsep}{4pt}
  \resizebox{\columnwidth}{!}{%
  \begin{tabular}{@{}p{0.60\columnwidth}c@{}}
    \toprule
    Metric & \% gain wrt control \\
    \midrule
    Employer High-relevance Rate ($\uparrow$)   & $+5.17\%$ ($+3.5\%$, $+6.8\%$) \\
    Employer Low-relevance Rate ($\downarrow$)   & $-27.31\%$ ($-35\%$, $-20\%$) \\
    Jobseeker High-relevance Rate ($\uparrow$)   & $+33.79\%$ ($+29.9\%$, $+37.7\%$) \\
    Jobseeker Low-relevance Rate ($\downarrow$)  & $-19.73\%$ ($-27\%$, $-13\%$) \\
    Apply Rate ($\uparrow$)                      & $+5.68\%$ ($0\%$, $+11.0\%$) \\
    Keep / (Keep+Remove) Rate ($\uparrow$) & $+7.07\%$ ($+5.0\%$, $+9.1\%$) \\
    \bottomrule
  \end{tabular}}
\end{table}

We report the relative lift of treatment over control
(Table~\ref{tab:online}), with 95\%
confidence intervals. We measure success along the same two relevance axes
used offline --- Employer Relevance and Jobseeker Relevance, each reported
as a high- and low-relevance rate. We also track two behavioral signals:
the keep rate, the fraction of surfaced candidates an employer
keeps rather than removes during review (a direct employer signal), and
the apply rate (a jobseeker signal).

The online numbers line up with what we saw offline and in simulation.
Swapping the additive heuristic for the learned single-token ranker
improved relevance for both employers and jobseekers, and it cut the share
of bad matches shown to employers by 27.3\%. The keep-rate result matters
most to us: it comes from employers actually keeping or removing surfaced
candidates, not from model-generated labels, so the 7.07\% gain is a sign
the improvement is real and not an artifact of evaluating against our own
labeler. Apply rate moved in the right direction as well, rising $5.68\%$,
which is directionally positive.

\section{System Architecture \& Engineering Implementation}

\begin{figure*}[t]
  \centering
  \includegraphics[width=\textwidth]{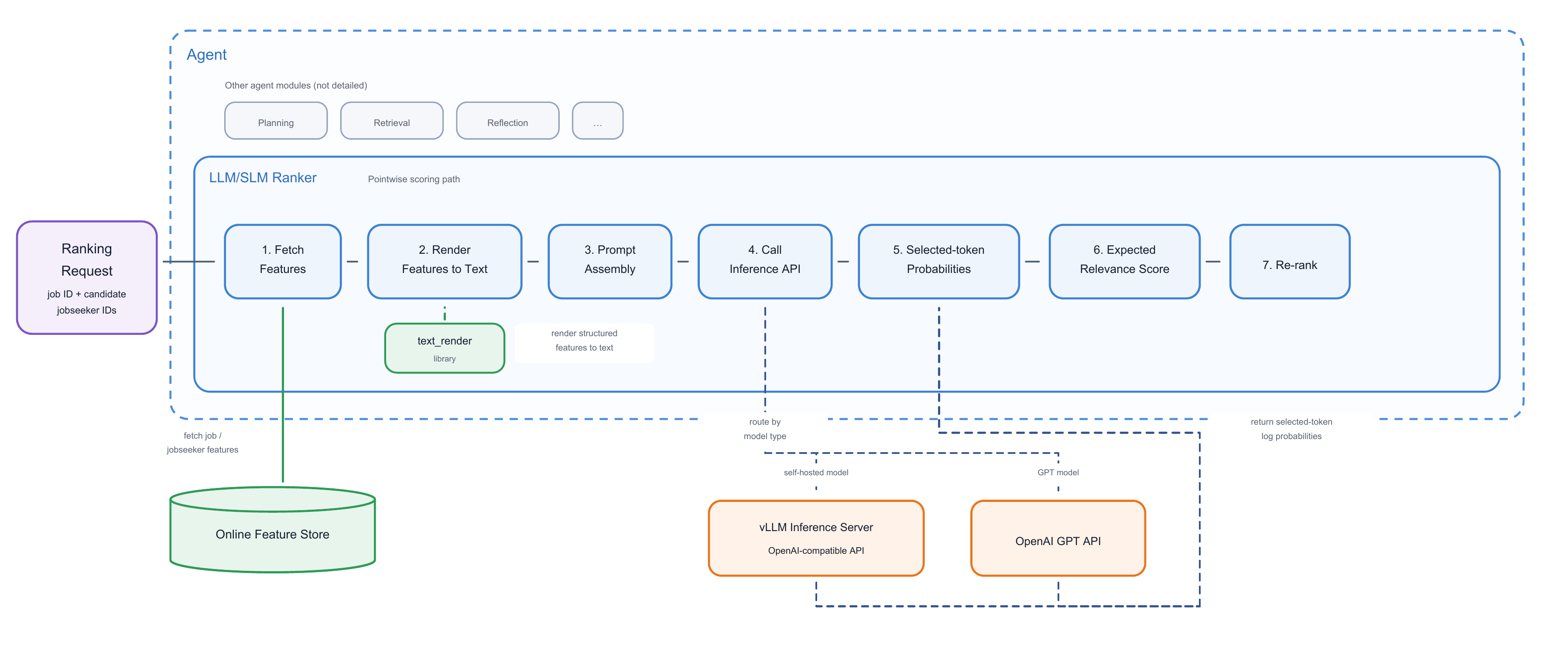}
  \caption{Serving architecture. At inference, candidate and job features
    are fetched from an online feature store and assembled into an
    LLM-ready prompt by a text-rendering step. Depending on the variant,
    scoring is served either by a vLLM inference server (Qwen3-8B with
    per-variant LoRA adapters) or by the OpenAI API; in both cases the
    model produces a distribution over the grade tokens $\{1,2,3,4,5\}$,
    from which we compute the expected value (ER) and rank candidates by
    it.}
  \label{fig:architecture}
\end{figure*}

The proposed ordinal ranker is deployed as the second-stage point-wise
scoring module within a multi-stage production pipeline, upstream of the
separate listwise reranking stage. Decoupling scoring from retrieval
confines heavy language model inference to a refined candidate subset,
keeping per-session latency low without sacrificing ranking quality. The
remainder of this section details the serving architecture
(Figure~\ref{fig:architecture}), data pipeline, latency optimizations,
production infrastructure, and operational monitoring.

\subsection{Data Pipeline and Consistency}

To mitigate training-serving skew, we employ a versioned rendering
configuration shared by training and serving. This ensures
identical feature hydration across offline training and online inference.
The rendered model input combines job-related information,
candidate-related information, matching constraints, and other
task-relevant contextual signals.

\subsection{Inference Latency Optimization}
\label{sec:latency}

To reduce inference latency, we implement three optimizations:
\begin{enumerate}
\item \textbf{Concurrent Scoring}: We utilize asynchronous I/O with
  shared connection pools to enable concurrent per-candidate scoring,
  avoiding request serialization.
\item \textbf{Feature-Fetch Overlapping}: Candidate metadata retrieval is
  executed in parallel with upstream matching logic, effectively masking
  network round-trip latency.
\item \textbf{Constrained Decoding}: By treating the task as a
  single-token classification, we remove the computational overhead of
  variable-length decoding and post-hoc output parsing.
\end{enumerate}

\subsection{Production Infrastructure}

The system is deployed on on-premise Kubernetes clusters as a vLLM-based
service, utilizing a shared foundation model (Qwen3-8B) with per-variant
LoRA adapters. This modular architecture supports multi-tenant serving,
where multiple ranking objectives share the same GPU memory footprint.

We optimize inference using fp8 precision for both weights and KV-cache,
alongside continuous batching and prefix caching to minimize
recomputation of the shared system prompt (Table~\ref{tab:infra}).
Capacity scales horizontally with demand to sustain throughput under
bursty, per-session fan-out patterns.

\begin{table}[ht]
  \centering
  \caption{vLLM serving and inference parameters.}
  \label{tab:infra}
  \footnotesize
  \setlength{\tabcolsep}{4pt}
  \begin{tabular}{@{}l p{0.58\columnwidth}@{}}
    \toprule
    Component & Configuration \\
    \midrule
    Base / adapter   & Qwen3-8B base + per-variant LoRA adapter \\
    Precision        & fp8 weights, fp8 KV cache \\
    Context length   & 16{,}384 tokens \\
    Runtime          & vLLM 0.22.0, self-hosted \\
    \bottomrule
  \end{tabular}
\end{table}

\subsection{Operational Monitoring}

We maintain stability through layered observability across three planes:
serving infrastructure, model quality, and service-level reliability.

On reliability, each scoring call is tagged with a mutually
exclusive outcome --- scored, no\_logprobs,
parse\_reject, or call\_error. Since any
non-scored call degrades gracefully to the deterministic heuristic, this
taxonomy separates infrastructure failure from silent quality loss. From
it we derive three production SLOs over a rolling 7-day window:
\emph{scoring availability} (non-error rate; target 99\%, observed
$\approx$99.95\%), \emph{usable-score rate} (the stricter fraction of
calls yielding a usable score; target 99\%, observed $\approx$99.8\%,
whose gap to availability isolates silent degradation), and
\emph{per-session stage latency} (p95 $\leq$ 8\,s, measured at the
per-session stage rather than the per-candidate scoring call of
Section~\ref{sec:latency}). Each
objective is guarded by multi-window burn-rate alerts, pairing a fast
window for acute outages with a slow window for gradual budget erosion,
and is defined as version-controlled SLO-as-code alongside per-variant
diagnostic monitors for triage.

Beyond standard infrastructure telemetry, we monitor the distribution of
ER scores and token logits to detect training/serving drift. Proposed
ranking models undergo a 48-hour shadow validation phase, auditing
performance against legacy baselines on live traffic prior to production
promotion.

\section{Analysis \& Discussion}
\label{sec:analysis}

\subsection{Fine-Grained Discrimination and Ordinal Consistency}

The pairwise breakdown in Figure~\ref{fig:pairwise} reveals a clear
pattern across the 1-to-5 ordinal labels, and it shows how well the
deployed model preserves numerical distance relative to the heuristic
baseline.

The model separates distant grades sharply --- 87.64\% pairwise accuracy
on the ``5 vs 1'' combination, which distinguishes highly qualified
candidates from entirely unqualified ones.

By contrast, distinguishing adjacent tiers such as ``5 vs 4'' and ``4 vs
3'' remains substantially harder, with accuracy in the range of roughly
61\%--64\%. This is consistent with the realities of recruitment, where
even human expert annotators face difficulty separating them. These
borderline qualifications occupy an inherently grey area, and the
residual error largely reflects this annotation ambiguity rather than a
failure of the model.

On adjacent combinations, the legacy heuristic performs close to random
guessing (approximately 51\%), whereas Qwen3-8B SFT sustains accuracy
above 61\%. The model thus preserves the ordering between neighboring
grades far better than the heuristic, rather than treating the labels as
unordered classes.

\subsection{Scaling Laws vs. Foundation Capacity Trade-offs}

Two figures bear on the trade-off between data efficiency and
foundation-model capacity: Figure~\ref{fig:data-scale} (GPT-4.1-mini at
20k vs.\ 50k) and Figure~\ref{fig:slm-vs-llm} (GPT-4.1-mini and Qwen3-8B
at a matched 50k).

As shown in Figure~\ref{fig:data-scale}, the gain from 20k to 50k samples
is small and quickly flattens for GPT-4.1-mini: that backbone reaches most
of its ranking quality by 20k labels. Data efficiency of this kind matters
for cold-start, low-traffic settings, where the large interaction logs that
deep neural rankers require are unavailable. It is, however, a property of
the backbone rather than of the task: Qwen3-8B is still improving well past
that point, gaining $+5.8\%$ to $+21.2\%$ from 50k to 200k pairs.

Figure~\ref{fig:slm-vs-llm} isolates the complementary limitation. With
training data held fixed at 50k, GPT-4.1-mini-ft leads Qwen3-8B-ft by
$+10.4\%$ to $+32.3\%$ across the seven metrics, so the difference is
attributable to the backbone rather than to supervision. Scaling Qwen3-8B
to 200k narrows but does not close it: with four times the training data
the deployed model remains $4.2\%$ to $8.4\%$ behind GPT-4.1-mini-ft at
50k. Beyond a certain point the bottleneck therefore shifts from data
volume to the representational capacity of the base model itself.

\subsection{Determinism: Variance Reduction and Latency Optimization}

\paragraph{Error cascades in long generation paths.}
Conventional two-stage approaches --- first emitting a complex JSON
structure or extracted reasoning, then producing a score --- require
longer generated output. Because the output is autoregressive, a small
probability deviation in an early token can be amplified in a cascade along
the decoding chain (an ``error cascade''), injecting variance into the
final score.

\paragraph{Constrained single-token decoding.}
Our approach instead extracts the logits at the very first generated token,
computes the expected value, and terminates the interaction immediately.
Because only a single token is decoded, there are no earlier tokens whose
small numerical fluctuations could cascade and be amplified into a large
deviation in the final score. In addition, cross-entropy fine-tuning drives
the model toward confident, low-entropy grade distributions
\cite{diversity_sft, confidence_penalty}, so the small floating-point
nondeterminism of GPU inference rarely changes the selected grade and only
negligibly perturbs the expected-value score. Together these remove the
sampling and generation-cascade sources of variance and greatly reduce the
model's per-call scoring variance, while also cutting decoding overhead.

\section{Summary and Future Work}

\subsection{Conclusion}

For cold-start and data-scarce recommendation settings, this work narrows
the gap between deep neural rankers that depend on
large click logs and hard-coded, linearly weighted heuristics. In their
place, we present an approach for fine-grained text alignment based
on SLM ordinal regression.

The proposed hybrid ordinal regression loss (MSE + CE), combined with
single-token expected-value scoring, preserves numerical monotonicity: on
adjacent grade pairs the deployed model stays above $61\%$ pairwise
accuracy where the heuristic is close to random. Alongside this, we
observe strong screening of high-value candidates and a sharp reduction
in low relevance rate ($-46.7\%$).

We show that a locally deployed SLM can
remove the generation variance of open-ended decoding and meet stringent
online latency requirements, while still drawing on the
world knowledge embedded in the underlying model.

The online experiment provides further evidence that the improvements
observed offline and in simulation are also realized in the production
environment. Improvements in model-assessed relevance were accompanied by
stronger employer interaction behavior, suggesting that the learned ranker
is surfacing quality candidates to employers.

Together, these results
offer a practical recipe for building talent-sourcing rankers under data
scarcity.

\subsection{Future Work}

\paragraph{From point-wise to list-wise fine-tuning.}
The current framework scores each candidate independently for a single
job. We plan to introduce list-wise objectives into the fine-tuning stage
so that the model learns not only to score individuals in isolation but
also to reason about the relative ordering among candidates within a job.

\paragraph{Further latency reduction.}
We will continue to optimize inference latency to push the system toward
even lower-cost, higher-throughput production serving.

\section*{Declaration on Generative AI}
During the preparation of this work, we used ChatGPT and Claude as writing and formatting aids, with every output reviewed and edited by us. Specifically, the tools were used to: check grammar and spelling; paraphrase and reword; improve writing style - sentence structure, word choice, and flow; create charts and diagrams from our data and experimental results; assist with citation management and latex formatting; and refine abstract. All research design, data, results, and conclusions are ours. No text or figures were used without our revision, and we take full responsibility for the publication's content.

\bibliography{references}

\begin{thebibliography}{14}
\expandafter\ifx\csname natexlab\endcsname\relax\def\natexlab#1{#1}\fi
\providecommand{\url}[1]{\texttt{#1}}
\providecommand{\href}[2]{#2}
\providecommand{\path}[1]{#1}
\providecommand{\DOIprefix}{doi:}
\providecommand{\ArXivprefix}{arXiv:}
\providecommand{\URLprefix}{URL: }
\providecommand{\Pubmedprefix}{pmid:}
\providecommand{\doi}[1]{\href{http://dx.doi.org/#1}{\path{#1}}}
\providecommand{\Pubmed}[1]{\href{pmid:#1}{\path{#1}}}
\providecommand{\bibinfo}[2]{#2}
\ifx\xfnm\relax \def\xfnm[#1]{\unskip,\space#1}\fi
\bibitem[{Dadaboyev et~al.(2025)Dadaboyev, Abdullayeva, Abbosova, Suleymenova,
  and Mamadjanova}]{aisourcing}
\bibinfo{author}{S.~M.~U. Dadaboyev}, \bibinfo{author}{J.~Abdullayeva},
  \bibinfo{author}{N.~Abbosova}, \bibinfo{author}{A.~Suleymenova},
  \bibinfo{author}{K.~Mamadjanova},
\newblock \bibinfo{title}{Role of artificial intelligence in employee
  recruitment: systematic review and future research directions},
\newblock \bibinfo{journal}{Discover Global Society}  (\bibinfo{year}{2025}).
\bibitem[{Han et~al.(2024)Han, Zhu, Hu, Qin, Zhao, and Zhu}]{llm4rec_hiring}
\bibinfo{author}{X.~Han}, \bibinfo{author}{C.~Zhu}, \bibinfo{author}{X.~Hu},
  \bibinfo{author}{C.~Qin}, \bibinfo{author}{X.~Zhao},
  \bibinfo{author}{H.~Zhu},
\newblock \bibinfo{title}{Adapting job recommendations to user preference drift
  with behavioral-semantic fusion learning},
\newblock in: \bibinfo{booktitle}{ACM SIGKDD}, \bibinfo{year}{2024}.
\bibitem[{Yu et~al.(2025)Yu, Xu, Xue, Zhang, Ma, and Yu}]{genai_recruitment}
\bibinfo{author}{X.~Yu}, \bibinfo{author}{R.~Xu}, \bibinfo{author}{C.~Xue},
  \bibinfo{author}{J.~Zhang}, \bibinfo{author}{X.~Ma}, \bibinfo{author}{Z.~Yu},
\newblock \bibinfo{title}{{ConFit} v2: improving resume-job matching using
  hypothetical resume embedding and runner-up hard-negative mining},
\newblock in: \bibinfo{booktitle}{Findings of ACL}, \bibinfo{year}{2025}.
\bibitem[{Qin et~al.(2018)Qin, Zhu, Xu, Zhu, Jiang, Chen, and
  Xiong}]{neural_ranker_data}
\bibinfo{author}{C.~Qin}, \bibinfo{author}{H.~Zhu}, \bibinfo{author}{T.~Xu},
  \bibinfo{author}{C.~Zhu}, \bibinfo{author}{L.~Jiang},
  \bibinfo{author}{E.~Chen}, \bibinfo{author}{H.~Xiong},
\newblock \bibinfo{title}{Enhancing person-job fit for talent recruitment: an
  ability-aware neural network approach},
\newblock in: \bibinfo{booktitle}{SIGIR}, \bibinfo{year}{2018}.
\bibitem[{Guo et~al.(2016)Guo, Alamudun, and Hammond}]{coldstart_heuristics}
\bibinfo{author}{S.~Guo}, \bibinfo{author}{F.~Alamudun},
  \bibinfo{author}{T.~Hammond},
\newblock \bibinfo{title}{R\'{e}su{M}atcher: a personalized r\'{e}sum\'{e}--job
  matching system},
\newblock \bibinfo{journal}{Expert Systems with Applications}
  (\bibinfo{year}{2016}).
\bibitem[{Liang et~al.(2025)Liang, Yang, Wang, Xu, Yu, and
  Shu}]{taxonomy_guided}
\bibinfo{author}{Y.~Liang}, \bibinfo{author}{L.~Yang},
  \bibinfo{author}{C.~Wang}, \bibinfo{author}{X.~Xu}, \bibinfo{author}{P.~S.
  Yu}, \bibinfo{author}{K.~Shu},
\newblock \bibinfo{title}{Taxonomy-guided zero-shot recommendations with
  {LLMs}},
\newblock in: \bibinfo{booktitle}{COLING}, \bibinfo{year}{2025}.
\bibitem[{Zhuang et~al.(2024)Zhuang, Qin, Hui, Wu, Yan, Wang, and
  Bendersky}]{finegrained_relevance}
\bibinfo{author}{H.~Zhuang}, \bibinfo{author}{Z.~Qin},
  \bibinfo{author}{K.~Hui}, \bibinfo{author}{J.~Wu}, \bibinfo{author}{L.~Yan},
  \bibinfo{author}{X.~Wang}, \bibinfo{author}{M.~Bendersky},
\newblock \bibinfo{title}{Beyond yes and no: Improving zero-shot {LLM} rankers
  via scoring fine-grained relevance labels},
\newblock in: \bibinfo{booktitle}{NAACL}, \bibinfo{year}{2024}.
\bibitem[{Ouyang et~al.(2025)Ouyang, Zhang, Harman, and Wang}]{llm_volatility}
\bibinfo{author}{S.~Ouyang}, \bibinfo{author}{J.~M. Zhang},
  \bibinfo{author}{M.~Harman}, \bibinfo{author}{M.~Wang},
\newblock \bibinfo{title}{An empirical study of the non-determinism of
  {ChatGPT} in code generation},
\newblock \bibinfo{journal}{ACM Transactions on Software Engineering and
  Methodology}  (\bibinfo{year}{2025}).
\bibitem[{Liu et~al.(2025)Liu, Shen, Shen, Yao, Kao, Xu, Arora, Zheng, Johnson,
  Hong, Wu, and Zhang}]{llm_query_understanding}
\bibinfo{author}{P.~Liu}, \bibinfo{author}{J.~Shen}, \bibinfo{author}{Q.~Shen},
  \bibinfo{author}{C.~Yao}, \bibinfo{author}{K.~Kao}, \bibinfo{author}{D.~Xu},
  \bibinfo{author}{R.~Arora}, \bibinfo{author}{B.~Zheng},
  \bibinfo{author}{C.~Johnson}, \bibinfo{author}{L.~Hong},
  \bibinfo{author}{J.~Wu}, \bibinfo{author}{W.~Zhang},
\newblock \bibinfo{title}{Powering job search at scale: {LLM}-enhanced query
  understanding in job matching systems},
\newblock in: \bibinfo{booktitle}{CIKM}, \bibinfo{year}{2025}.
\bibitem[{Guo et~al.(2025)Guo, Li, Zhuang, Luo, Li, Yan, Zhu, and
  Zhang}]{aspect_matching}
\bibinfo{author}{F.~Guo}, \bibinfo{author}{W.~Li}, \bibinfo{author}{H.~Zhuang},
  \bibinfo{author}{Y.~Luo}, \bibinfo{author}{Y.~Li}, \bibinfo{author}{L.~Yan},
  \bibinfo{author}{Q.~Zhu}, \bibinfo{author}{Y.~Zhang},
\newblock \bibinfo{title}{{MCRanker}: generating diverse criteria on-the-fly to
  improve pointwise {LLM} rankers},
\newblock in: \bibinfo{booktitle}{WSDM}, \bibinfo{year}{2025}.
\bibitem[{Shi et~al.(2025)Shi, Ma, Liang, Diao, Ma, and
  Vosoughi}]{judge_position_bias}
\bibinfo{author}{L.~Shi}, \bibinfo{author}{C.~Ma}, \bibinfo{author}{W.~Liang},
  \bibinfo{author}{X.~Diao}, \bibinfo{author}{W.~Ma},
  \bibinfo{author}{S.~Vosoughi},
\newblock \bibinfo{title}{Judging the judges: A systematic study of position
  bias in {LLM}-as-a-judge},
\newblock in: \bibinfo{booktitle}{IJCNLP-AACL}, \bibinfo{year}{2025}.
\bibitem[{Pei et~al.(2024)Pei, Pang, Cai, Sengupta, and
  Toshniwal}]{ordinal_relevance}
\bibinfo{author}{Y.~Pei}, \bibinfo{author}{Y.~W. Pang},
  \bibinfo{author}{W.~Cai}, \bibinfo{author}{N.~Sengupta},
  \bibinfo{author}{D.~Toshniwal},
\newblock \bibinfo{title}{Leveraging {LLM} generated labels to reduce bad
  matches in job recommendations},
\newblock in: \bibinfo{booktitle}{RecSys}, \bibinfo{year}{2024}.
\bibitem[{Li et~al.(2025)Li, Chen, Xu, Qin, Xiao, Luo, and Sun}]{diversity_sft}
\bibinfo{author}{Z.~Li}, \bibinfo{author}{C.~Chen}, \bibinfo{author}{T.~Xu},
  \bibinfo{author}{Z.~Qin}, \bibinfo{author}{J.~Xiao}, \bibinfo{author}{Z.-Q.
  Luo}, \bibinfo{author}{R.~Sun},
\newblock \bibinfo{title}{Preserving diversity in supervised fine-tuning of
  large language models},
\newblock in: \bibinfo{booktitle}{ICLR}, \bibinfo{year}{2025}.
\bibitem[{Pereyra et~al.(2017)Pereyra, Tucker, Chorowski, Kaiser, and
  Hinton}]{confidence_penalty}
\bibinfo{author}{G.~Pereyra}, \bibinfo{author}{G.~Tucker},
  \bibinfo{author}{J.~Chorowski}, \bibinfo{author}{{\L}.~Kaiser},
  \bibinfo{author}{G.~Hinton},
\newblock \bibinfo{title}{Regularizing neural networks by penalizing confident
  output distributions},
\newblock in: \bibinfo{booktitle}{ICLR Workshop}, \bibinfo{year}{2017}.

\end{thebibliography}

\end{document}